\documentclass[%
 reprint,
superscriptaddress,
 amsmath,amssymb,
 aps,
 pra,
]{revtex4-1}

\usepackage{graphicx}
\usepackage{dcolumn}
\usepackage{bm}
\usepackage{hyperref}
\usepackage{color}

\newcommand{\VRF}[1]{\textcolor{black}{ #1}}

\begin{document}


\title{A modal description of paraxial structured light propagation}

\author{Hend Sroor}
\affiliation{School of Optical-Electrical and Computer Engineering, University of Shanghai for Science and Technology, Shanghai, China}
\author{Chane Moodley}
\affiliation{School of Physics, University of the Witwatersrand, Johannesburg, South Africa}
\author{Valeria Rodr\'iguez-Fajardo}
\affiliation{School of Physics, University of the Witwatersrand, Johannesburg, South Africa}
\author{Qiwen Zhan}
\affiliation{School of Optical-Electrical and Computer Engineering, University of Shanghai for Science and Technology, Shanghai, China}
\author{Andrew Forbes}
\affiliation{School of Physics, University of the Witwatersrand, Johannesburg, South Africa}

\date{\today}

\begin{abstract}
\noindent Here we outline a description of paraxial light propagation from a modal perspective.  By decomposing the initial transverse field into a spatial basis whose elements have known and analytical propagation characteristics, we are able to analytically propagate any desired field, making the calculation fast and easy.  By selecting a basis other than that of planes waves, we overcome the problem of numerical artefacts in the angular spectrum approach and at the same time are able to offer an intuitive understanding for why certain classes of fields propagate as they do.  We outline the concept theoretically, compare it to the numerical angular spectrum approach, and confirm its veracity experimentally using a range of instructive examples.  We believe that this modal approach to propagating light will be a useful addition to toolbox for propagating optical fields.  
\end{abstract}

\maketitle

\section{Introduction}
\noindent Our understanding of the propagation of light has refined over the centuries, starting with geometric approaches that have their foundation in concepts outlined more than 400 years ago, through to a wave description that was given a firm theoretical footing nearly 200 years later \cite{young:light}, the two married through notions of stationary phase, least action and path interference \cite{gitin2013huygens}.  From Maxwell onwards we have been able to calculate the propagation of arbitrary optical fields directly from the wave equation, finding exotic scalar solutions that include accelerating light \cite{efremidis2019airy}, non-diffracting light \cite{mazilu2010light}, the eigenmodes of free-space in various co-ordinate systems \cite{gutierrez2005helmholtz} and vectorial light \cite{zhan2009cylindrical,rosales2018review}, generally referred to today as structured light \cite{roadmap,forbes2020structured,forbes2021structured}.  The classes and their properties are more commonly grouped and understood using geometrical \cite{padgett1999poincare,Holleczek2011,Milione2012higher,alonso2017ray,gutierrez2019modal} and operator \cite{Stoler:81,dennis2019gaussian} perspectives, shedding a deeper understanding on their commonality.  In the context of laser beams, statistical tools applied to scalar fields has revealed the commonality in behaviour of all classes of beams \cite{siegman1991defining}, for example, that their second moment widths and divergences follow the same propagation rule as Gaussian beams but with adjusted beam quality factors \cite{siegman1993defining}, while a quantum toolkit has been successfully applied to vectorial light to quantify its degree of vectorness \cite{McLaren2015,Ndagano2016} and many other parameters \cite{qian2015shifting} by entanglement measures, exploiting parallels between non-separability in vector beams and quantum entangled states \cite{forbes2019classically,konrad2019quantum,Eberly2016}. 

But how to calculate the propagation \VRF{of} these exotic fields? The standard textbook approach is to use the angular spectrum method by decomposing the fields into a basis of plane waves \cite{goodman2005introduction}. Its numerical nature means that it suffers from lack of physical insight into the nature of the propagation, although it does \VRF{lead} itself to easy implementation both on a computer and in the laboratory for ``digital'' propagation of paraxial light \cite{Schulze2012C}.

\begin{figure*}[ht!]
\centering\includegraphics[width=\linewidth]{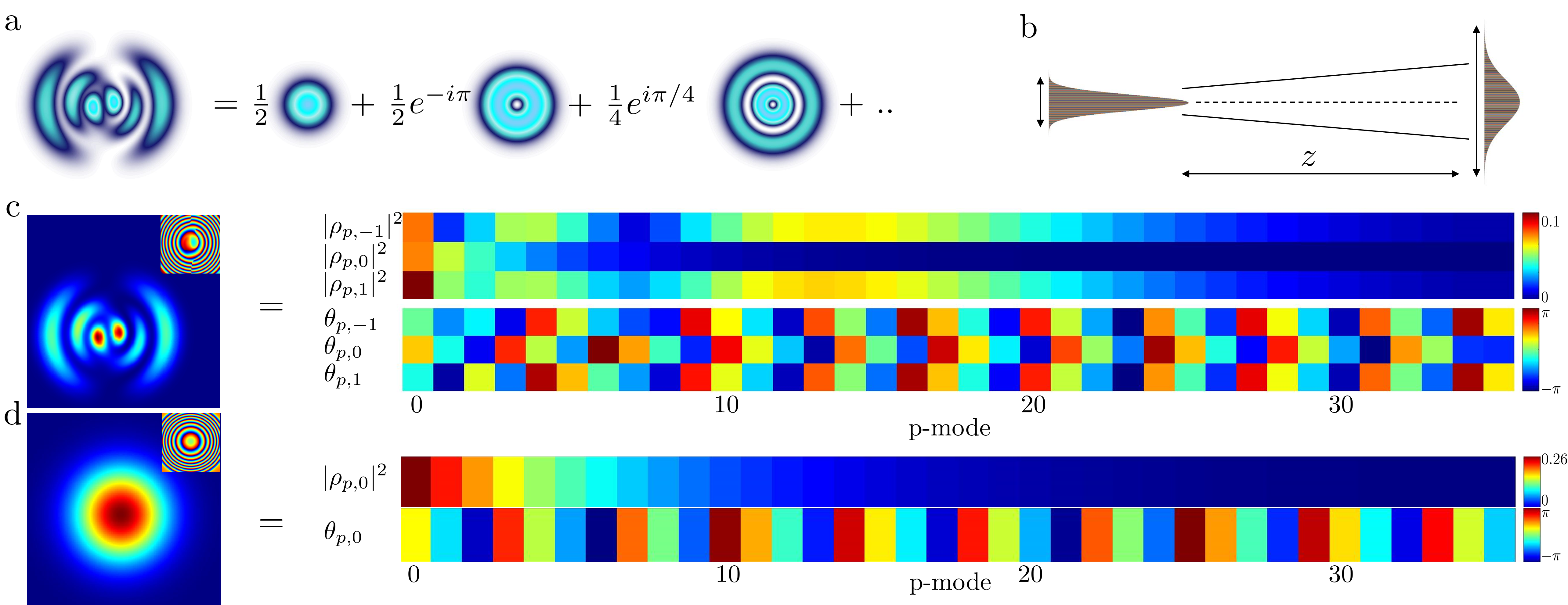}
\caption{(a) An arbitrary mode, shown on the \VRF{L}HS, can be decomposed into a sum of eigenmodes with complex weightings\VRF{, on the RHS}. (b) Each eigenmode on the \VRF{R}HS has an analytical $z$-dependent propagation, whose sum returns the propagation of the arbitrary mode. Two examples of the decomposition are shown in (c) and (d), where the desired fields are expressed as $\ell$ and $p$ Laguerre-Gaussian modes with the amplitudes $|\rho_{p,\ell}|^2$(top rows) and phases $\theta_{p,\ell}$(bottom rows) shown as false colour weights.}
\label{fig:concept}
\end{figure*}

In this work we outline a modal approach to the propagation of arbitrary optical fields, shown graphically in Fig.~\ref{fig:concept}.  We decompose an initial field at the plane $z=0$ into an appropriate basis with a known $z$-dependent propagation function.  Because each basis element in the decomposition can be propagated analytically, so can the entire initial field which may not have any known analytical propagation rule.  We use our approach to offer an intuitive explanation for some well-known propagation properties, including the propagation invariance of certain classes of modes, why lenses focus light, and why there is a ``far field'' where the light's propagation remains shape invariant.  To illustrate the ease of implementation and accuracy of the approach, we compare it to the numerical angular spectrum approach, itself a type of modal analysis, showing excellent agreement, and then validate the method by experiment.  Although we have restricted our examples to the ubiquitous scalar paraxial transverse spatial modes of light for brevity, it should be clear that it works equally well for vectorial light fields by applying the decomposition twice, once for each vector component.  Similarly, it can be adapted to the time domain by using a one dimensional basis with known dispersion in certain media.  We believe that this approach is powerful, intuitive and will be a useful resource in both research and teaching laboratories alike.
 
\section{Spatial modes as a basis}
Consider that we wish to perform a modal expansion of an initial field, $u(x,y,z=0)$ into some orthonormal basis $\psi_{i}(x,y,z=0)$ where 

\begin{equation}
u (x,y,z=0) = \sum_{i} c_{i} \psi_{i} (x,y,z=0)\,.
\label{modal0}
\end{equation}

To find the unknown coefficients $c_{i}$ one performs a modal decomposition, which can be done numerically or optically \cite{pinnell2020modal}, to return 

\begin{equation}
c_{i} = \iint u (x,y,0)\psi^*_{i} (x,y,z=0) dA\,.
\label{modaldecomp}
\end{equation}

If the basis has a known propagation rule, i.e.,  $\psi_{i}(x,y,z=0) \overset{z}{\longrightarrow} \psi_{i}(x,y,z)$, then since each basis element on the RHS has a known $z$ dependence, we can find the propagation of the LHS by

\begin{equation}
u (x,y,z) = \sum_{i} c_{i} \psi_{i} (x,y,z)\,.
\label{modal0}
\end{equation}

We will show that this simple expansion is powerful both computationally and intuitively.  Since the transverse plane is two-dimensional, examples of appropriate bases would be the Hermite-Gaussian (HG) or Laguerre-Gaussian (LG) beams (to name but two), given by
\begin{equation}
u (x,y,z) = \sum_{n,m} c_{n,m} \text{HG}_{n,m}(x,y,z) = \sum_{p,\ell} c_{p,\ell} \text{LG}_{p,\ell}(r,\phi,z)\,,
\label{modalz}
\end{equation}

We will use these two bases in the remainder of this report although any orthonormal basis will do. For example, in the Hermite-Gaussian basis we have 

\begin{multline}
\text{HG}_{n,m}(x,y,z) = \sqrt{I_0} \text{H}_{n} \left( \frac{\sqrt{2} x}{w(z)} \right) \text{H}_{m} \left( \frac{\sqrt{2} y}{w(z)} \right) \\
\times \exp \left( -\frac{x^2 +y^2}{w^2(z)} \right) \exp ( i \eta(x,y,z) )\,,
\end{multline}

\noindent where
\begin{align}
\eta (x,y,z) &= -kz - k \frac{x^2 + y^2}{2R(z)} + (m+n+1) \arctan \left( \frac{z}{z_R}\right)\,, \nonumber \\
z_R &= \frac{\pi w_0^2}{\lambda}\,, \nonumber \\
w(z) &= w_0 \sqrt{1 + \frac{z^2}{z^2_R}}\,, \nonumber \\
R(z) &= z \left( 1 + \frac{z^2_R}{z^2} \right)^{-1}\,, \nonumber
\end{align}

\noindent $w_0$ is the embedded Gaussian beam size at $z=0$ and $I_0$ is found by normalising the power to $P$, to return

\begin{equation}
I_0 = 2 \mu_0c \frac{w^2_0}{w^2(z)} \frac{P}{\pi w^2(z) n! m! 2^{m+n-1}}\,.
\end{equation}

We can change to the Laguerre-Gaussian basis

\begin{multline}\label{LG}
\text{LG}_{p,\ell}(r,\phi) = \sqrt{I_0} \text{L}_p^{\ell} \left( \frac{2r^2}{w^2(z)}\right) \left( \frac{\sqrt{2}r}{w(z)}\right)^{|\ell|}
\\ \times \exp \left( -\frac{r^2}{w^2(z)} \right) \exp ( i \eta(r,\phi,z) )\,,
\end{multline}

\noindent where

\begin{align}
    \eta (r,\phi,z) &= -kz - \frac{kr^2}{2R(z)} -\ell \phi + (2p+|\ell|+1) \arctan \left( \frac{z}{z_R}\right)\,, \nonumber \\
I_0 &= 2 \mu_0c\frac{2P p!}{\pi w^2(z) (p + |\ell|)!)}\,, \nonumber
\end{align}

\noindent and all other terms share the functional form of the HG modes.  These two examples are pertinent since they encompass both Cartesian and cylindrical symmetries.

\section{Modal propagation}
 
To understand what influences the propagation dynamics, we note that the modal expansion is considered complete, so that the modal powers add up to the total power of the field, which we will set to $P=1$ for convenience.  Using the HG modes as an example, this means that $\sum_{n,m} |c_{n,m}|^2 = 1$.  However, the modal weightings are in general complex numbers, $c_{n,m} = \rho_{n,m} \exp{i \theta_{n,m}}$, with modal amplitudes ($\rho_{n,m}$) and phases ($\theta_{n,m}$). The initial field can be viewed as the interference of many HG modes.  But in free-space there is no coupling between the HG modes, that is, there is no power exchange where one HG mode gains power at the expense of other, so the modal powers remain invariant.  Likewise, the initial modal phases are constants that do not change with distance, but they appear to be altered by the Gouy phase change that is both mode and distance dependent, $\eta_{n,m}(z) \propto (m+n+1) \arctan( \frac{z}{z_R})$.  The phase change $\Theta_{n,m}(z) = \theta_{n,m} + \eta_{n,m}(z)$ therefore holds all the information on how the propagation of any arbitrary mode (on the LHS \VRF{of Eq.~\ref{modalz}}) will evolve, since this term causes the various HG modes to interfere either constructively or destructively, altering with distance.  In the modal perspective it is this interference that determines how optical modes propagate.  In the case of the HG and LG bases, all modes have exactly the same radius of curvature, independent of basis indices.  This crucial fact eliminates all spatial dependence from the phase, making it only a function of $z$.  

What do we gain from such an expansion? There are at least two advantages: (1) the propagation becomes computationally simple since one need only perform a modal decomposition once, on the initial field, and thereafter only analytical propagation of each basis element is performed; (2) the propagation becomes more intuitive, which we illustrate with the examples to follow.
\\
\\
\noindent \textbf{Eigenmodes of free space I:} why are some optical fields propagation invariant in free-space? For example, the HG and LG modes do not change their intensity profile (shape) during propagation and instead only slowly diverge in size - we will call this ``propagation invariant''. In the modal propagation scenario, if the optical field in question is $u(x,y) = \text{HG}_{n,m}(x,y)$ then an expansion following Eq.~\ref{modalz} becomes

\begin{equation}
u (x,y,z) = \sum_{n,m} c_{n,m} \text{HG}_{n,m}(x,y,z) = \text{HG}_{n,m}(x,y,z)\,.
\end{equation}
\noindent since $|c_{n,m}|^2 = 1$.  As there is only one eigenmode on the RHS, there can be no interference and hence no change to the optical field during propagation.  Thus, if the initial field can be written in some coordinate system where it appears as a basis element, then in that symmetry it will be propagation invariant.
\\
\\
\noindent \textbf{Eigenmodes of free space II:} if the interference between terms in the expansion never changes, then the field to be propagated must also be propagation invariant.  This can happen when the mode number of each basis element is identical, i.e., $N_{n,m} = n + m + 1$ or $N_{p,\ell} = 2p + |\ell| + 1$.  The modal propagation approach predicts that there must be an infinite set of propagation invariant modes from appropriate superpositions of basis elements, and not just the trivial case shown above.  For example, an initial petal-like mode of $u(r,\phi) = 2 \cos (6 \phi)$ can be written as

\begin{equation}
u(r,\phi) = 2 \cos (6\phi) \overset{z}{\longrightarrow} \text{LG}_{0,3}(r,\phi,z) + \text{LG}_{0,-3}(r,\phi,z)\,.
\end{equation}
Here the phase change of each mode is the same with $z$, so that the interference pattern remains unaltered with distance, as does the field itself.  In Cartesian coordinates

\begin{equation}
u(x,y) = \VRF{\frac{1}{\sqrt{2}}} \text{HG}_{5,3}(r,\phi,0) + \VRF{\frac{1}{\sqrt{2}}} \text{HG}_{1,7}(r,\phi,0)\,,
\end{equation}
\noindent will also be propagation invariant during propagation, and so on.
\\
\\
\noindent \textbf{General modes in free space:} A counter-example is the general case when the field to be propagated is arbitrary.  In this case the phase change of each basis element brings them in and out of constructive interference with the others, so that the final mode itself must also change during propagation.  An example of this is a flat-top beam, expressed as 
\begin{eqnarray}
u(x,y) = \exp\bigg(-\frac{\big(x^{2n} + y^{2n}\big)}{w_0^{2n}}\bigg),
\label{eq:ft}
\end{eqnarray}
\noindent where $n$ represents the order of the supergaussian. Such beams change dramatically in shape during propagation \cite{gori1994flattened}.
\\
\\
\noindent \textbf{The far-field:} Why do all optical fields converge to a ``far field'' pattern after multiples of the Rayleigh range ($z_R$) and thereafter do not alter in structural form with distance (other than a scale change)? In the modal propagation interpretation it is because the modal phases themselves converge to $\Theta_{n,m}(z) \rightarrow \pi (n+m+1)/2 + \theta_{n,m}$ at $z >> z_R$, i.e., a constant.  Thus the relative phase of each basis element remains unchanged after this distance, and therefore so does the interference between the modes.  If the interference does not change, then neither can the field that is propagating.  This is why all fields eventually become ``propagation invariant''.  The converse is equally true.  If $z << z_R$ then the functional dependence of the phases is negligible so that $\Theta_{n,m}(z) \rightarrow  \theta_{n,m} $.  Here again there is no change in interference with distance and so the field must be close to ``propagation invariant'', but virtue of a large Rayleigh range.
\\
\\
\begin{figure}
\centering\includegraphics[width=\linewidth]{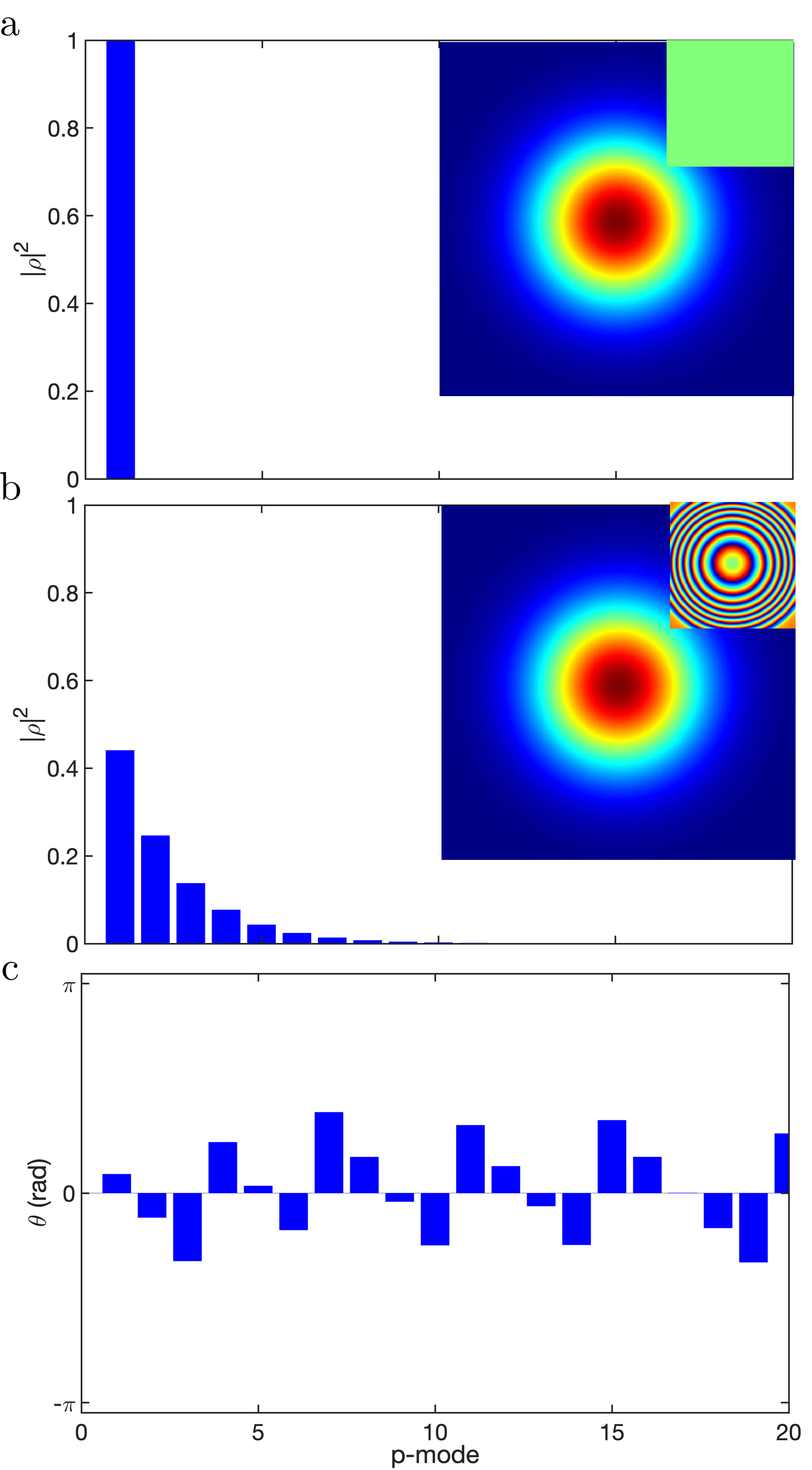}
\caption{(a) When a Gaussian beam with a planar wavefront is decomposed into the LG basis at $z=0$, only the $p=0$ (Gaussian term) has a non-zero weighting. But if a Gaussian beam with curvature phase is decomposed into the same basis, many radial ($p$) modes appear, each with a modal phase which changes linearly with $p$, shown  in (b) and (c), respectively. The insets shown the Gaussian intensity and phase as illustrative false colour plots.}
\label{fig:gauss}
\end{figure}
\noindent \textbf{Why lenses focus:} An intriguing aspect of a modal decomposition is that the choice of basis and basis parameters determines the number of modes in the expansion, as shown graphically in Fig.~\ref{fig:gauss}.  For example, a Gaussian beam with curvature, $\text{LG}_{0,0} (z=0) \exp(ikr^2/2f)$, when expressed into the LG basis with no curvature (planar wavefronts with $R=0$), returns a superposition of many radial modes.  
However, a Gaussian beam with curvature is equivalent to passing a planar wavefront Gaussian through a lens.  This then suggests that the focussing of lenses can be given a modal explanation: the interference of many radial modes gives rise to the lensing action, predicting that the beam should converge to a spot (or diverge if the sign is reversed). How to make sense of this and see it in the ``math''?  The complex expansion coefficients for this special case can be calculated analytically and found to be

\begin{equation}
c_{p} =  |c_p| \exp \left( i p \arctan (k w_0^2/4f) \right)\,,
\end{equation}
\noindent with
\begin{equation}
|c_p| = 4f(k w_0^2)^p (16 f^2 + k^2 w_0^4)^\frac{p}{2}\,,
\end{equation}
\noindent where a constant modal phase of $\exp ( i \arctan (-k w_0^2/4f))$ across all modes has been dropped (since it is the relative phase that matters). We see that the phase scales \textit{linearly} with mode order: $\theta (p) = p \times \arctan (k w_0^2/4f) $.  However, we know that a lens phase function scales quadratically with distance.  To see the connection, we approximate the Laguerre-Gaussian beam as an oscillating cosine function (by linking the Laguerre-Gaussian beam to its Bessel equivalent \cite{mendoza2015laguerre} and the Bessel equivalent to the cosine function \cite{litvin2008bessel}), finding $\cos ( 2 \sqrt{2p+1} r/w_0 - \pi/4)$, so that for large $p$ the $p^\text{th}$ ring will be approximately located at $r_p^2 \approx p (w_0 \pi/16) ^2$. Since $r^2 \propto p$ and $\theta \propto p$ we have that $\theta \propto r^2$, as needed for a lensing action.  It is instructive to compare this situation to the binary ring construction of a lens, by finding those radii where the light arrives in phase at the desired focal spot a distance $f$ from the $r=0$ position.  This is easily calculated to be $r_n^2 \approx 2 n \lambda f$: the zones scale linearly with ring number (diffraction order) $n$. In the modal case, these rings are constructed from carefully crafted phase variations, done automatically by the modal decomposition, while the amplitudes account for the distribution of the light.  

With the intuitive benefits now highlighted, we move on to demonstrate the easy of implementation, both numerically and experimentally.

\section{Experimental and Numerical Validation}

\begin{figure}[t]
\centering\includegraphics[width=\linewidth]{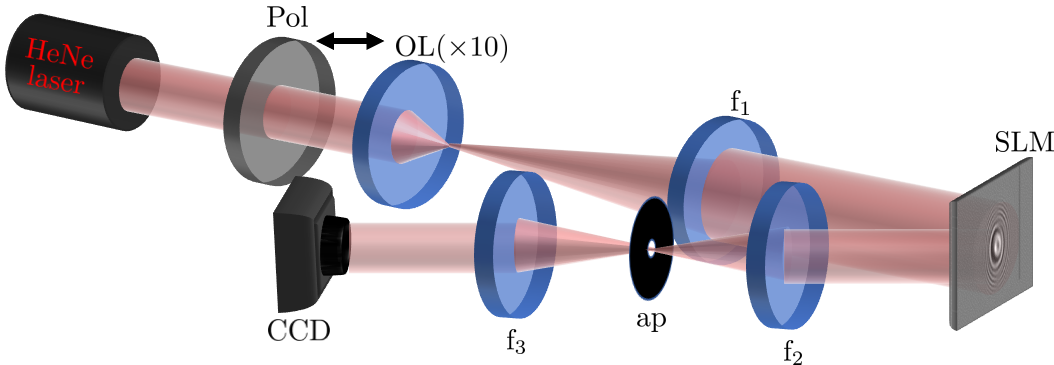}
\caption{Illustration of the experimental set-up, where a spatial light modulator (SLM) was used to digitally create the desired fields to be tested.}
\label{fig:setup}
\end{figure}

\begin{figure}[h!]
\centering\includegraphics[width=\linewidth]{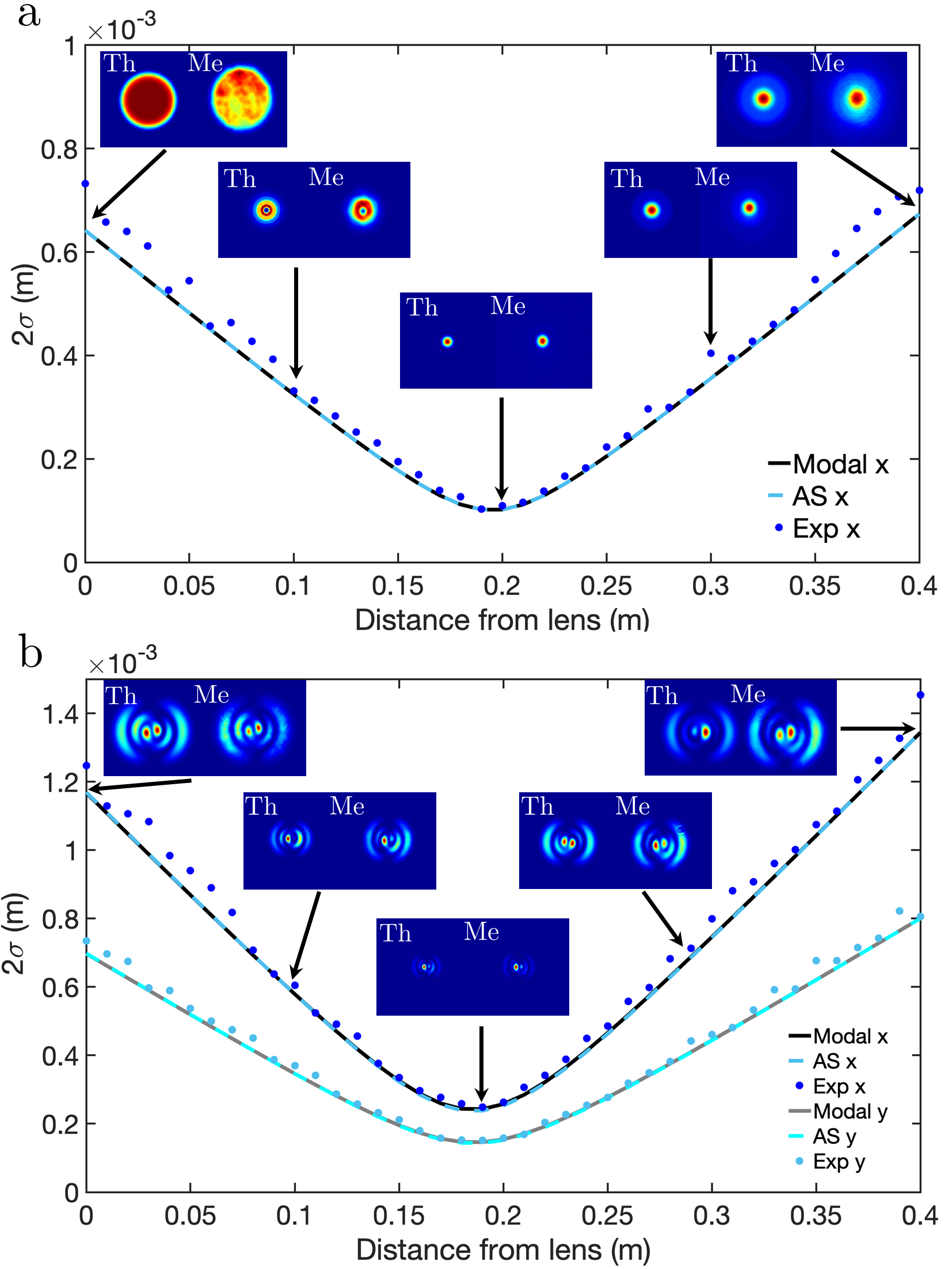}
\caption{Second moment beam widths in the $x-$ and $y-$axis as a function of propagation distance ($z$) for the (a) flat-top and (b) exotic beam, comparing experimentally measured widths (Exp) to those calculated by the angular spectrum (AS) and modal (Modal) approaches. The insets show the measured intensities (Me) and the theoretical (Th) intensities from the modal approach.}
\label{fig:width}
\end{figure}

\noindent To validate of this concept, we perform an experiment and a numerical comparison to the common angular spectrum approach.  Our experimental set-up, shown in Fig.~\ref{fig:setup}, makes use of a visible laser beam and a Spatial Light Modulator (SLM). A Gaussian beam from a HeNe laser ($\lambda=633$ nm) was passed through a polarizer (Pol), oriented for horizontal polarization, before \VRF{being} expanded \VRF{by} an 10$\times$ objective lens (OL) and then collimated by a $f$ = 300 mm lens to overfill \VRF{the} SLM (HoloEye, PLUTO-VIS, with 1920 $\times$ 1080 pixels of pitch 8 um and calibrated for a 2$\pi$ phase shift at $\lambda = 633$ nm). The SLM  was encoded with an appropriate computer generated hologram to create the desired field to be tested, often requiring complex amplitude modulation \cite{Forbes2016, SPIEbook}. The desired mode was imaged by lenses $f_2$=$f_3$ = 300 mm, with the aperture (ap) at the Fourier plane used to remove unwanted diffraction orders. A Point Grey Firefly camera was used to measure the beam profiles from the image plane ($z=0$) as a function of $z$ by moving the camera on a rail. The second moment width of the beam at each position was calculated from the captured images.  To measure the far-field and to observe the beams passing through their waist planes, we employed a digital lens of focal length $f$ programmed on the SLM rather than a physical lens.
\begin{figure}[h!]
\centering\includegraphics[width=\linewidth]{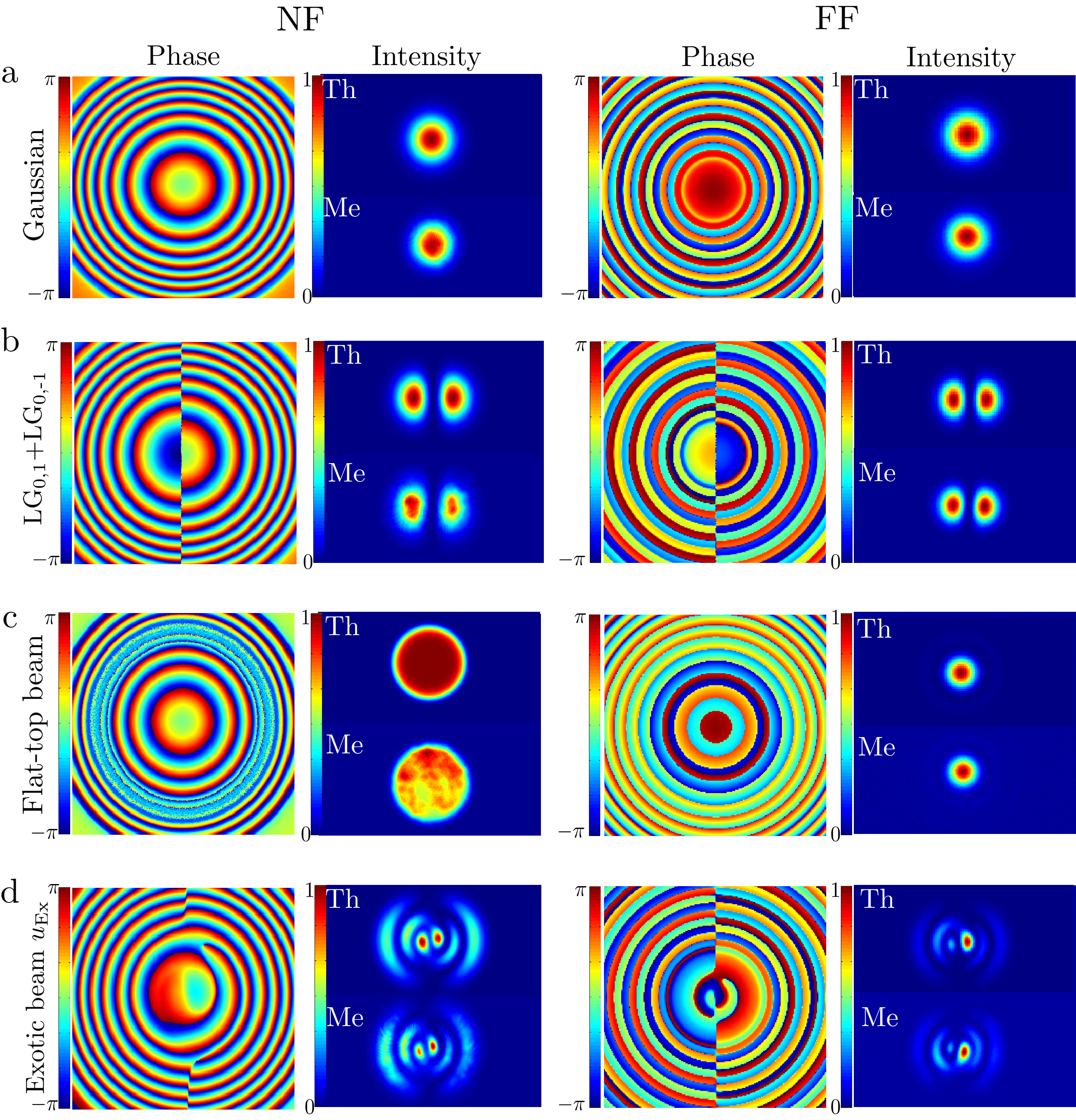}
\caption{Near-field (NF) and far-field (FF) results for the four test cases using the modal propagation approach, showing intensity and phase. An additional curvature $\text{exp}(ikr^2/2f)$ with $f=200$ mm is been added to total phase of each test case in order to measure the far-field plane. Each row of the figure shows from left to right the reconstructed phase of the beam at $z = 0$, a comparison of the intensity profiles of the numerically calculated (Th) and experimentally measured (Me) beam intensity at $z=0$, the reconstructed phase at $z = f$ mm, a comparison of the profiles of the numerically calculated (Th) and experimentally measured (Me) beam intensity at $z= f$ mm, respectively.}
\label{fig:NF-FF}
\end{figure}
We selected four test cases to cover a range of possibilities: (1) a Gaussian beam, (2) a superposition of LG$_{0,\pm1}$ beams, (3) a flat-top beam with $n=10$ and (4) the exotic mode similar to Fig.~\ref{fig:concept} (a), given by

\begin{multline}
    u_{\text{Ex}}= 0.5e^{i\pi}\text{LG}_{0,1}+0.25e^{i\frac{\pi}{4}}\text{LG}_{1,-1}+e^{-i\frac{\pi}{2}}\text{LG}_{1,2} \\ +0.5\text{LG}_{0,0}+0.25e^{i\frac{\pi}{4}}\text{LG}_{1,1}+e^{-i\frac{\pi}{2}}\text{LG}_{2,-1}\,.
\end{multline}
The results of these tests are shown in Figures \ref{fig:NF-FF} and \ref{fig:width}.

Results in Fig.~\ref{fig:NF-FF} shows good agreement between the simulated beam profiles using our modal propagation approach (Th) and the measured profiles (Me), both in the near field ($z=0$) and far field ($z=f$). The results are consistent with the predictions: that some modes will be propagation invariant and others not. The fact that the results are corroborated experimentally in both the near and far fields confirms that it must work for all propagation distances. 
To quantify the agreement, we measure beam images from $z=0$ to $z = 400$ mm and calculate the second moment beam radius in the two orthogonal axis, with the results for the flat-top and exotic beams shown in Figure~\ref{fig:width}.  We overlay on the results a calculation from the  traditional angular spectrum (AS) approach, which was applied here by using the 2D Fast-Fourier transform (FFT) algorithm of the optical field and then the inverse 2D FFT to obtain superposition of all propagated planar wave components in the observation plane. It is evident that there is excellent agreement between calculated (AS and modal) and measured (Exp) results. 

\section{Conclusion}
The modal approach to propagating arbitrary forms of structured light has the advantages of being analytic, computationally simple, and offers physical insights into the propagation dynamics of classes of modes.  Here we have outlined the approach, used it to offer an intuitive understanding of paraxial light propagation, validated it against the traditional angular spectrum method and confirmed it experimentally.  Although this test was limited to scalar light, it is self-evident that it will work for vectorial light too by applying the approach to both polarization components of the field.
\section*{Funding}
The authors would like to thank the 
National Science Foundation of China (NSFC) grant, grant agreement no 92050202, China Post-doctoral Science Foundation grant, grant agreement no 238691,and the CSIR-IBS (South Africa) bursary scheme.

\section*{Disclosures}
The authors declare no conflicts of interest.

\bibliography{progress}
\end{document}